\documentclass[aps,prl,reprint]{revtex4-1}

\usepackage{graphicx}
\usepackage{dcolumn}
\usepackage{bm}

\usepackage[utf8]{inputenc}
\usepackage[ngerman,english]{babel}
\usepackage{lmodern,hfoldsty,charter}
\usepackage{color}
\usepackage{amsmath}
\usepackage{amssymb}
\usepackage{units}

\begin{document}
 \title{Experimental realisation of $\mathcal{PT}$-symmetric flat bands}
 \date{\today}

 \author{Tobias Biesenthal}
 \author{Mark Kremer}
 \author{Matthias Heinrich}
 \author{Alexander Szameit}
 \email{alexander.szameit@uni-rostock.de}
 \affiliation{Institut für Physik, Universit\"at Rostock, Albert-Einstein-Str. 23, 18059 Rostock, Germany}

\begin{abstract}
\noindent The capability to temporarily arrest the propagation of optical signals is one of the main challenges hampering the ever more widespread use of light in rapid long-distance transmission as well as all-optical on-chip signal processing or computations. To this end, flat-band structures are of particular interest, since their hallmark compact eigenstates do not only allow for the localization of wave packets, but importantly also protect their transverse profile from deterioration without the need for additional diffraction management. In this work, we experimentally demonstrate that, far from being a nuisance to be compensated, judiciously tailored loss distributions can in fact be the key ingredient in synthesizing such flat bands in non-Hermitian environments. We probe their emergence in the vicinity of an exceptional point and directly observe the associated compact localised modes that can be excited at arbitrary positions of the periodic lattice.

\end{abstract}

\maketitle

Shaping and steering the flow of light remains one of the core objectives in optics, particularly in the realm of integrated photonics. Recent years have seen dramatic progress in methods that employ structural modifications of the monolithic host medium to achieve this goal. The perhaps best known example are photonic crystals \cite{PC_Joannopoulos,PCF_Knight,PCF_Russell}, where the strong periodic refractive index modulation represented by certain hole patterns gives rise to gaps in the band structure that suppress light propagation at certain wavelengths and angles of incidence. Similarly, waveguide arrays with much lower index contrast are likewise characterized by band structures that govern the discrete transverse dynamics \cite{Discretizing_Christodoulides}. In this context, the task of slowing down or entirely arresting the displacement or broadening of wave packets is inextricably linked to the concept of flat bands, which have been explored in a variety of different settings, in one-dimensional \cite{flat_band_rhombic, sawtooth, flat_band_rhombic_2, flat_band_stub, graphene_ribbons} as well as in two-dimensional settings \cite{Lieb_lattice,loc_flat_band_lieb,band_collapse_graphene,Strain_Graphene_Landau}.

At the same time, non-Hermitian physics, spearheaded by its representatives, parity-time ($\mathcal{PT}$) symmetry \cite{PT_Bender_Boettcher} and exceptional points \cite{berry_non_herm,weiss_ex_p,Guo_PRL}, provides new insights into the interplay of the real and imaginary parts of complex potentials, and allows these quantities to be exploited as dynamic degrees of freedom instead of static global parameters used merely to compensate each other. Photonics is particularly suited to reap the benefits of these ongoing research efforts, since complex-valued potentials naturally translate to particular distributions of refractive index, gain and loss \cite{deme_OL,deme_PRL,deme_solitons}. To date, $\mathcal{PT}$-symmetry and exceptional points were demonstrated experimentally in various settings, ranging from pairs of coupled waveguides \cite{PT_in_optics,Guo_PRL} to complex photonic systems with one and two spatial dimensions \cite{nat_comm_toni, nat_mater_weimann, nat_rotter, 2D_PT}, coupled fiber loops \cite{peschel_1d,peschel_1d_bloch,peschel_bloch_pt} and even microring lasers arrangements \cite{HodaeiScience}.

Despite its fundamental importance for controlling the flow of light, recent technological advances in $\mathcal{PT}$-symmetric photonics have not yet enabled the  realization of flat bands in $\mathcal{PT}$-symmetric structures. Here, we experimentally demonstrate that flat bands and their associated compact localized states can indeed be established at the exceptional point of $\mathcal{PT}$-symmetric lattices. By introducing precisely tailored losses, we are able to observe the signature diffraction-less long distance propagation in entirely passive arrays of evanescently coupled waveguides.


\begin{figure}
	\center \includegraphics[width=\linewidth]{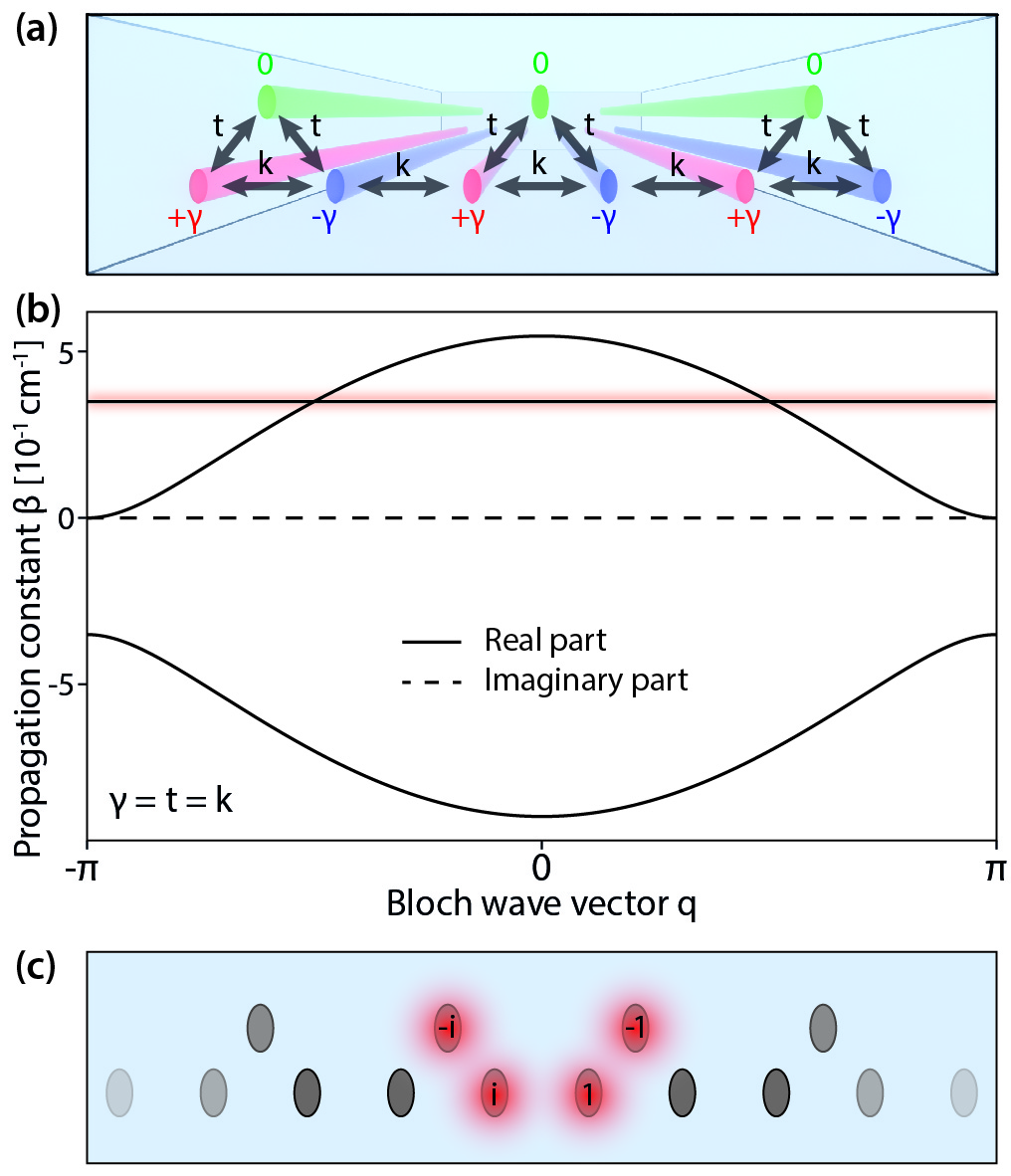}
 	\caption{(a) Schematic of the tripartite $\mathcal{PT}$-symmetric photonic lattice under consideration. The unit cell is comprised of three sites, one with neutral imaginary part (green), one with loss $-\gamma$ (blue) and one with gain $+\gamma$ (red). The coefficients $k$ and $t$ characterize the strengths of envanescent coupling.
(b) Resulting dispersion relation at the exceptional point of the structure for $\gamma=t=k$. Note that, if present, a global loss factor manifests itself as a shifted imaginary part across the entire Brillouin zone.
(c) Field distribution of the compact localized states associated with the flat band of this structure. The trapezoidal arrangement comprises four appropriately phase-shifted waveguides with identical amplitudes and partially overlaps with two adjacent unit cells.}
\end{figure}

The unit cell of the tripartite tight-binding lattice under consideration consists of a triangular arrangement of waveguides with identical real parts of their on-site potential. Figure 1(a) illustrates how these unit cells are arranged in a quasi-one-dimensional chain in which sites $a,c$ with gain (red, imaginary part $+\gamma$) and loss (blue, imaginary part $-\gamma$) are coupled with a coefficient $k$ in an alternating fashion, whereas the central site $b$ (green) of each unit cell has a "neutral" imaginary part, i.e. the average of the gain/loss sites, and interacts with both of them with the coefficient $t$. This arrangement can be described by the discrete Schrödinger equation $i\frac{d}{dz}\psi_n=\hat{H}_q\psi_n$, where $z$ denotes the propagation coordinate, $\psi_n=(a_n,b_n,c_n)^T$ is the three-component wave function describing the field amplitude in unit cell $n$, and the Hamiltonian $\hat{H}$ reads as
\begin{align}\label{eq:Hamiltonian}
	\hat{H}_q=\left(
		\begin{smallmatrix}
			{0} & {-t} & {-k-k e^{-iq}}\\
			{-t} & {-i \gamma} & {-t}\\
			{-k-k e^{iq}} & {-t} & {-2 i \gamma}
		\end{smallmatrix}
	\right)\,.
\end{align}
As shown by Ramezani et al. \cite{PT_flat_band_Ramezani}, this arrangement undergoes its phase transition from unbroken to broken $\mathcal{PT}$-symmetry as the contrast of the imaginary part is increased to the threshold value of $\gamma_{\mathcal{PT}}=t\sqrt{2-t^2/k^2}$. The two upper bands gradually flatten and approach each other with increasing $\gamma$, until they finally fuse at the exceptional point. The resulting flat band extends across the entire Brillouin zone (see Fig. 1(b)) and features a propagation constant of $\beta_0=t^2/k$. From Eq.~\eqref{eq:Hamiltonian}, one then finds the corresponding eigenmodes to have the form $\psi_q=\left(1,\nicefrac{-t}{\beta_0}(1+\xi),\xi\right)^T$ where $\xi=[\beta_0-\nicefrac{t^2}{\beta_0}-i\gamma]/[\nicefrac{t^2}{\beta_0}-k (1+e^{-iq})]$. In the spatial domain, these compact eigenstates involve contributions from two adjacent unit cells, e.g.  $\Psi_{n} = (0, -t/\beta_0 \xi^*, 1/\xi^*)^T$ and $\Psi_{n+1} = (1, -t/\beta_0, 0)^T$.
Choosing the two coupling strengths to be equal ($k=t$) dramatically simplifies the structure of this mode to feature identical amplitudes and only phase shifts between all involved sites:
\begin{equation}\label{eq:eigenstate}
	\Psi_{n} = (0,-i,i)^T \quad \textrm{and} \quad \Psi_{n+1} = (1,-1,0)^T.
\end{equation}
The corresponding trapezoidal wave packet and the relative phases between its respective lattice sites are illustrated in Fig.1(c).


\begin{figure}
 \center
 \includegraphics[width=\linewidth]{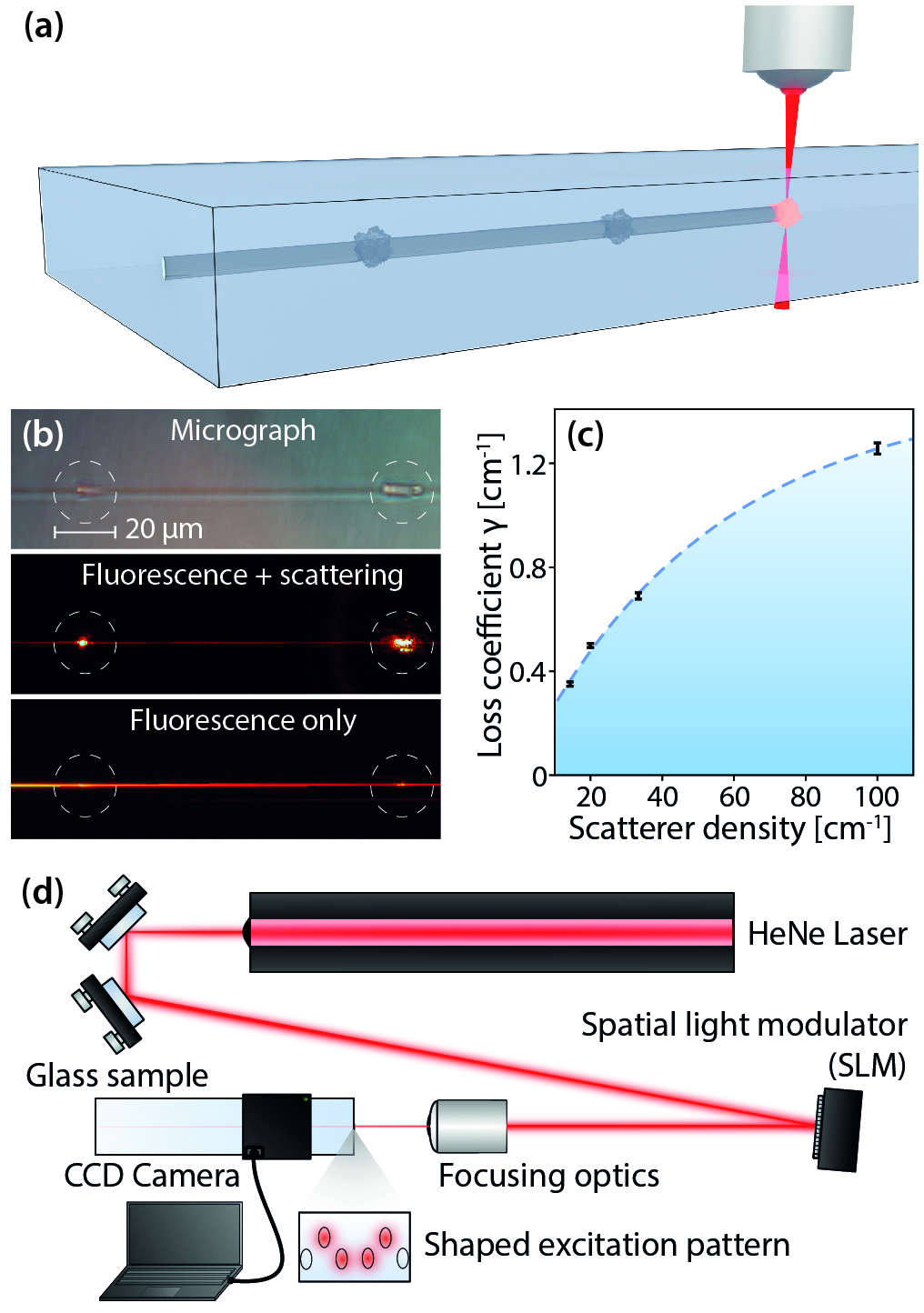}
 \caption{(a) The desired amounts of additional losses were implemented by introducing an appropriate concentration of scattering centers during the inscription process.
 (b) Whereas each scattering dot expels a small fraction of the propagating light, fluorescence imaging of the propagation pattern remains unaffected due to the spectral separation of the propagating and scattered light from the fluorescence signal.
 Top: Phase contrast micrograph of a typical waveguide with two subsequent scattering dots. Middle: Fluorescence micrograph without spectral filtering. Bottom: Spectrally filtered fluorescence micrograph.
 (c) Seamless tunability of the effective loss coefficient via the concentration of scattering dots.
 (d) A spatial light modulator (SLM) was used to synthesize the amplitude- and phase distribution for the excitation of flat band states.}
\end{figure}

\begin{figure}[h!]
 \center
 \includegraphics[width=.95\linewidth]{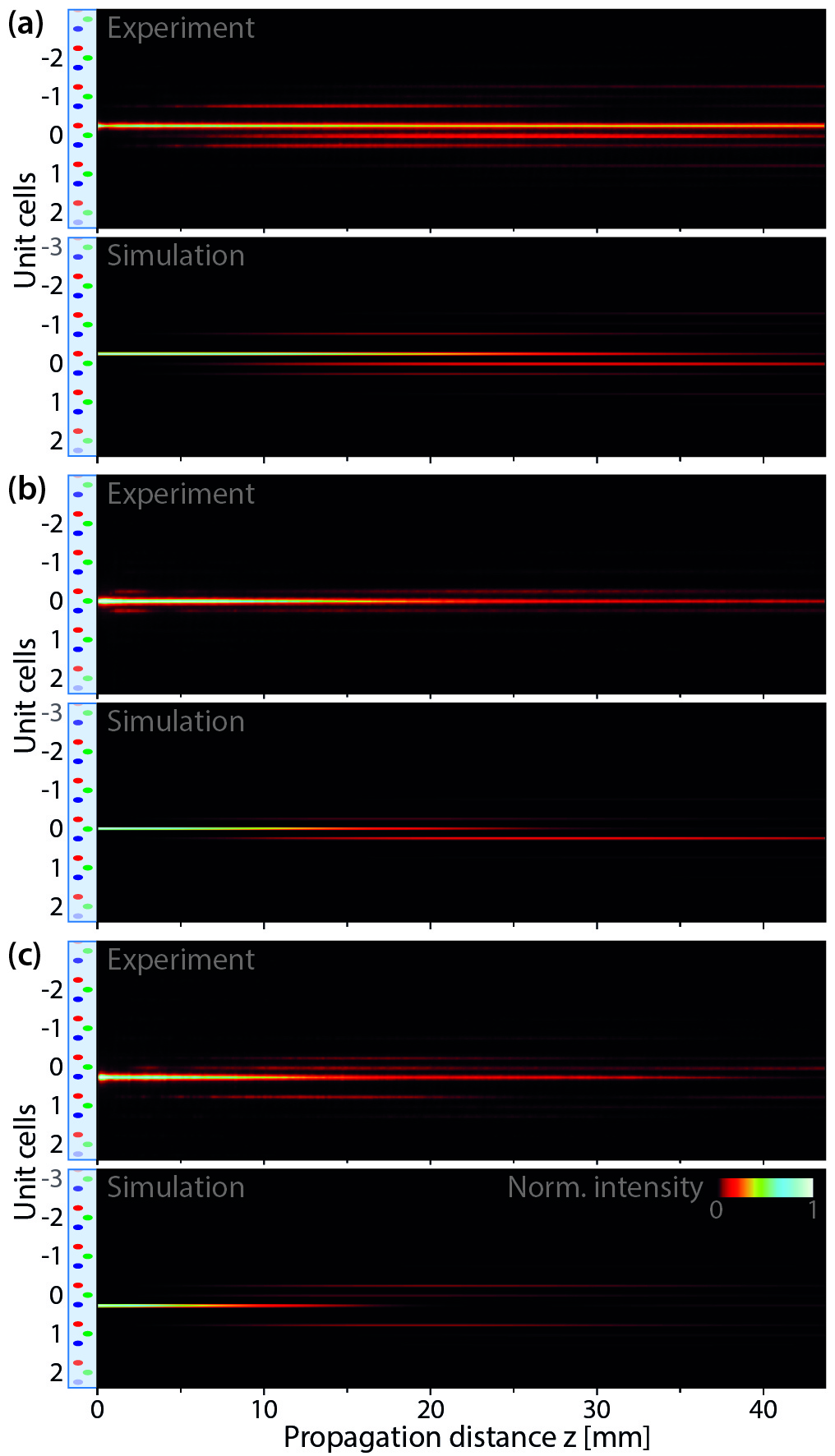}
 \caption{Intensity propagation dynamics resulting from single-site excitations at the three different waveguides of the unit cell: (a) "gain" site $a$, (b) "neutral" site $b$, and (c) "lossy" site $c$. The lattice parameters in the experimental system were set to $k=t=\gamma=0.3 \textrm{cm}^{-1}$. In all cases, the exponential decay of the propagating wave packet is due to the entirely passive implementation of the lattice, with imaginary parts $0,-\gamma$ and $-2\gamma$, which were also used in the numerical simulations. The top panel depicts the observed patterns, whereas the lower one shows the numerically calculated behavior.}
\end{figure}

 A challenge in implementing this structure in an experimental setting is the need for multiple, precisely tuned values of loss and gain. In conventional $\mathcal{PT}$-symmetric settings with only two levels of the imaginary part of the on-site potential, it is sufficient to realize their difference, as the exponential decay factor associated with a global imaginary offset faithfully preserves the propagation dynamics of the system \cite{Quasi_PT_symmetry}. While we made use of this latter fact to avoid the need for optical amplification by shifting the respective lattice sites from $\gamma$, $0$ and $-\gamma$ to $0$, $-\gamma$ and $-2\gamma$, the system at hand still necessitates a precise control over the amount of loss in each lattice site. To this end, we utilized the femtosecond laser direct writing technique \cite{Discrete_optics_Szameit_Nolte} and inscribe photonic lattices in accordance with the geometry sketched in Fig. 1(a). Losses were introduced by means of microscopic scattering centers \cite{2D_PT} that were generated by a brief pause of the longitudinal motion during the inscription process (see Fig. 2(a)). As shown in Fig. 2(b), whereas each individual scatterer only expels a small fraction of the propagating light (typically $\lesssim4\%$), changes to their concentration (i.e. spacing along the propagation direction) and scattering strength (index contrast and physical size, both of which increase with longer dwelling times) allowed us to continuously tune the overall propagation loss of the modified waveguide (see Fig. 2(c)). Notably, the point-like character of the scattering centers readily allows for such lossy waveguides to be arranged in arbitrary non-planar and even 2D configurations, leaving the real part of their effective refractive index virtually unchanged. At the same time, potential resonant re-capture effects of expelled light between subsequent scatterers in the same waveguide or in adjacent channels of the lattice \cite{modulated_wg_Eichelkraut} are minimized.

In order to probe the dynamics of the fabricated lattice, we used a Helium-Neon laser and synthesized different excitation patterns with a spatial light modulator (see Fig. 2(b)). These were subsequently projected onto the sample front facet, allowing us to observe the corresponding propagation patterns with waveguide fluorescence microscopy \cite{Nonlinear_refr_index_fslwg,quasi_incoherent_prop}. The spectral separation of the injected light ($633\,\mathrm{nm}$) and the fluorescence signal ($\approx 650\,\mathrm{nm}$) allows for quantitative intensity measurements of the propagating wave packet even in the presence of considerable damping. In addition to blocking scattered light with an edge pass filter, we employed Fourier filtering to reduce background noise without distorting the actual propagation dynamics to be observed.
Single-waveguide excitations populate the entire band structure and therefore yield strongly diffracting wave packets, regardless of which site of the unit cell is excited. This is shown in detail in Fig. 3. The case where light is injected into a "gain" waveguide, that is, a waveguide with minimal loss is shown in Fig. 3(a), in the experiment (top) and the simulation (bottom). In Fig. 3(b) a "neutral" site, that is, a site with intermediate loss, is excited, showing again a broadening of the wavepacket in experiment (top) and simulation (bottom). A broadening of the wave packet is also visible when a "loss" site with maximal loss is excited (see Fig. 3(c), with experiment (top) and simulation (bottom)).

The situation changes completely when the excitation pattern matches the amplitude- and phase distribution of the trapezoidal flat-band states. In this case, broadening of the wavepacket is visibly suppressed, as shown in Fig. 4(a) in experiment (top) and simulation (bottom). In order to quantify the stark difference between those two types of excitations, we numerically extracted the relative broadening as measured in terms of the second moment of the intensity distributions $\sigma^2(z)$. Finally, a normalization to their respective initial values $\sigma^2(0)$ allows for an easier comparison in the face of the intrinsically different diffraction rates associated with wider wave packets. In close agreement with the predicted behavior, Figure 4(b) shows how the eigenmode excitation is virtually free of broadening in the observed range of propagation, whereas the single-site excitations continuously diffract, and thereby dramatically increase in width.

\begin{figure}[h!]
 \center
 \includegraphics[width=.95\linewidth]{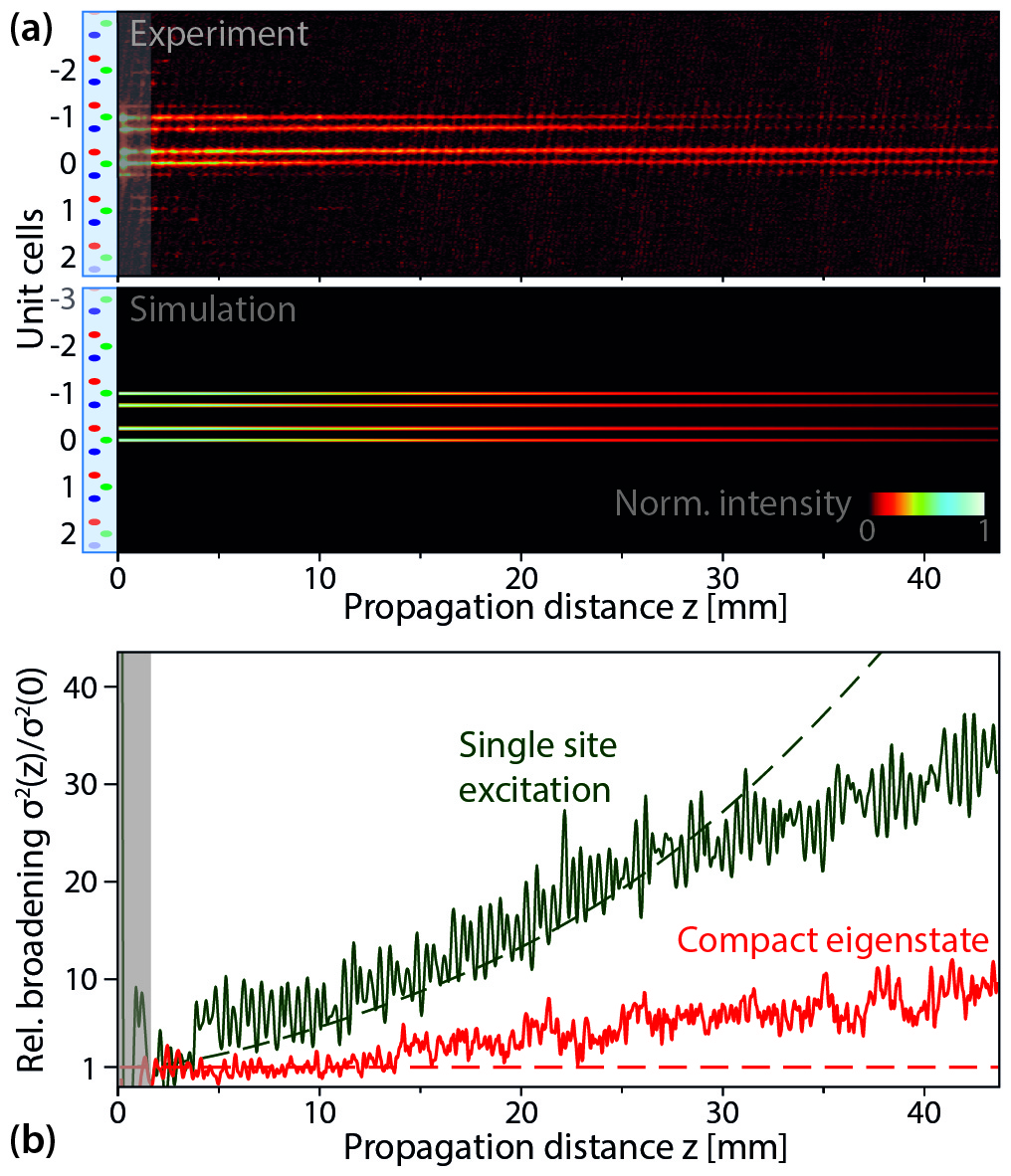}
 \caption{(a) Observed diffraction-free propagation of an excited trapezoidal compact eigenstate (top) and corresponding numerically calculated behavior (bottom).
 (b) Relative broadening of the eigenstate excitation compared to the single-site excitation (data from Fig. 3(a)). Shown are the width $\sigma^2(z)$ of the propagating wave packets, normalized with respect to their respective initial widths $\sigma^2(0)$. The experimental data from the first $2\,\mathrm{mm}$ was excluded from this evaluation since the signal in this region (shaded gray) is dominated by fluorescence excited by stray light traversing the lattice, and not the actual wave packet propagating within the guides themselves. As reference, the dashed graphs represent the numerically calculated behavior in both cases.}
\end{figure}

In our work, we created flat bands in $\mathcal{PT}$-symmetric optical systems and observed their characteristic compact localised eigenmodes. With this first demonstration, using laser-written photonic lattices with judiciously tailored loss distributions, we show that even in scenarios aiming to arrest the propagation and diffractive broadening of optical signals, losses are not necessarily detrimental, and can, in fact, serve as key ingredient in achieving the desired photonic flat band response in non-Hermitian environments.

\section{Acknowledgments}
AS gratefully acknowledge financial support from the Deutsche Forschungsgemeinschaft (grants SZ 276/9-1, SZ 276/19-1, SZ 276/20-1) and the Alfried Krupp von Bohlen und Halbach Foundation. The authors would also like to thank C. Otto for preparing the high-quality fused silica samples used in all experiments presented here.

\end{document}